\documentclass[aps,prl,twocolumn,groupedaddress,showpacs]{revtex4}

\usepackage{graphicx}
\usepackage{dcolumn}
\usepackage{bm}
\usepackage{subfigure}
\usepackage{amsmath}
\usepackage{amssymb}

\begin{document}


\title{Spin-1/2 Optical Lattice Clock}

\author{N.~D.~Lemke\footnote{also at University of Colorado, Boulder, CO, 80309}}
\author{A.~D.~Ludlow}
\author{Z.~W.~Barber\footnote{present address: Spectrum Lab, Montana State University, Bozeman, MT 59717}}
\author{T.~M.~Fortier}
\author{S.~A.~Diddams}
\author{Y.~Jiang\footnote{also at East China Normal University, Shanghai, 200062, China}}
\author{S.~R.~Jefferts}
\author{T.~P.~Heavner}
\author{T.~E.~Parker}
\author{C.~W.~Oates\footnote{electronic address: oates@boulder.nist.gov}}

\affiliation{National Institute of Standards and Technology \footnote{Contribution of U.S. government; not subject to copyright.}\\325
Broadway, Boulder, CO 80305}

\date{\today}

\begin{abstract}
We experimentally investigate an optical clock based on $^{171}$Yb ($I=1/2$) atoms confined in an optical lattice.  We have evaluated all known frequency shifts to the clock transition, including a density-dependent collision shift, with a fractional uncertainty of $3.4~\times~10^{-16}$, limited principally by uncertainty in the blackbody radiation Stark shift.  We measured the absolute clock transition frequency relative to the NIST-F1 Cs fountain clock and find the frequency to be 518~295~836~590~865.2(0.7)~Hz.  
\end{abstract}

\maketitle

Optical frequency standards based on the $^1S_0\rightarrow\!^3P_0$ transition in alkaline earth-(like) atoms confined in an optical lattice \cite{Katori03} have recently demonstrated performances approaching or exceeding those of the best microwave primary frequency standards \cite{Ludlow08,Poli08,Baillard07,Akatsuka08}.  Although these clocks have not yet achieved the same level of uncertainty as the best single trapped-ion clocks \cite{Rosenband08}, they have additional promise in that the large number of atoms simultaneously interrogated (typically tens of thousands) should produce lower quantum projection noise, eventually leading to increased clock stability \cite{Itano93,Hollberg01}.  Early work on lattice clocks has principally focused on strontium \cite{Katori03,Ludlow08,Baillard07}, and its only abundant fermionic isotope, $^{87}$Sr, has nuclear spin $I=9/2$.  Here we present an experimental study of a lattice clock employing the simplest fermionic structure available, a spin-1/2 system.  Such a system also lends itself well to other investigations, including quantum information applications \cite{Hayes07,Gorshkov08}.

Our previous work on Yb lattice clocks employed a bosonic isotope, $^{174}$Yb ($I=0$) \cite{Barber06,Poli08}, which offers the simplicity of an effective two-level system.  A spin-1/2 system has only slightly more complicated level structure, and reduces the need for careful calibration of field-induced shifts that may limit the accuracy of even isotope-based clocks \cite{Taichenachev06}.  Furthermore, a spin-1/2 system offers two advantages over higher angular momentum systems: optical pumping of atoms to the desired $m_F$ ground state is more straightforward, and the tensor component of the lattice light shift is absent \cite{Angel68}.  Although any clock based on a fermionic isotope needs to have its first-order Zeeman sensitivity removed, this is readily accomplished by averaging over two transitions with opposite dependence (see Fig.~1a) \cite{Ludlow08,Baillard07,Takamoto06}. 

Benefitting from the simplicity of the spin-1/2 system, our first realization of a $^{171}$Yb lattice clock has demonstrated results competitive with those of the best lattice clocks to date.  By comparing the Yb clock with other optical and microwave frequency standards via a femtosecond-laser comb \cite{Fortier06, Stalnaker07}, we have canceled or calibrated all known systematic effects with a total uncertainty of 3.4~$\times~10^{-16}$ and measured the absolute frequency with an uncertainty of 1.4~$\times~10^{-15}$.  We have observed a  density-dependent collision shift that persists despite the use of ultracold, spin-polarized fermions.  Additionally, in anticipation of higher-dimensional lattice geometries that will suppress non-zero density shifts, we have measured the vector polarizability of the clock transition to be $\alpha_v \approx -$0.08 a.u.  

\begin{figure}
\includegraphics[width=8cm]{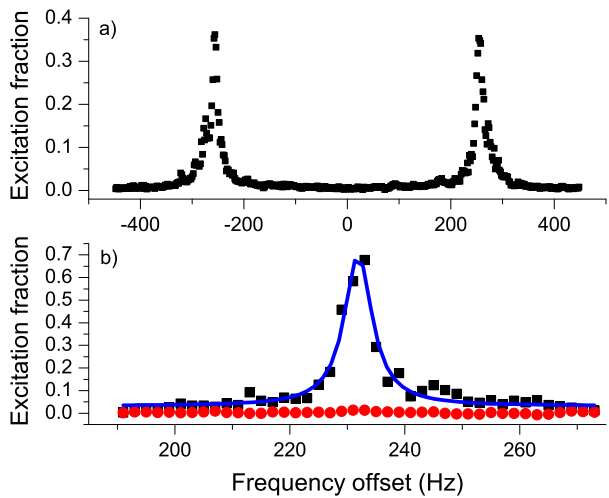}
\caption{(a) Measured $^1S_0\rightarrow\!^3P_0$ spectrum with an externally applied magnetic field of B=0.11 mT and $\pi$-polarized probe light.  The first-order Zeeman effect splits the two resonances by 210 Hz/G from center.  (b) Zoomed in scan of one of the $\pi$ transitions, with atoms pumped to (squares) and away from (circles) the $m_F=-1/2$ spin state.  A Lorentzian fit (solid line) gives a Fourier-limited FWHM of 5.5 Hz (for 160 ms probe time), which represents a quality factor of $10^{14}$, with over 60 $\%$ contrast.  Both (a) and (b) were averaged five times to suppress clock laser frequency noise.\label{spectrum}}
\end{figure}

The experimental sequence is similar to that we have described previously \cite{Poli08,Barber06}.  $^{171}$Yb atoms are trapped and cooled, first on the strong $^1S_0\leftrightarrow\!^1P_1$ transition at 399~nm, and second on the narrower $^1S_0\leftrightarrow\!^3P_1$ transition at 556~nm.  The atoms are then loaded into a one-dimensional, horizontally oriented optical lattice of 500~$E_r$ depth (recoil energy $E_r/k_B$ = 100~nK).  Approximately $3~\times~10^4$ atoms with a temperature of 15~$\mu$K are captured in several hundred lattice sites, leading to an estimated density of $\rho_0 =10^{11}$/cm$^3$.  To reduce line-pulling effects from other transitions, and in hopes of reducing density shifts through the Pauli exclusion principle \cite{Campbell09,Akatsuka08}, we pump the population to a single ground state, alternating experimental cycles between $m_F=\pm1/2$, before exciting the clock transition.  This state polarization is accomplished by optically pumping on the $^1S_0\leftrightarrow\!^3P_1$ transition at 556~nm, with which we transfer the atoms to a single ground state with 99 $\%$ efficiency (Fig.~1b).

During spectroscopy of the $^1S_0\rightarrow\!^3P_0$ clock transition at 578~nm, we apply a small magnetic field of 0.11~mT.  The lattice polarization and probe polarization are aligned with the vertical magnetic field, and a $\pi$-pulse resonant with the clock transition is applied for 80~ms, yielding a Fourier-limited lineshape of 10~Hz. The 578~nm probe light is pre-stabilized to a high finesse, environmentally isolated cavity, which produces a fractional instability of $\approx2~\times~10^{-15}$ in 1 s with the drift held below 1~Hz/s.  We detect the fraction of excited atoms by applying a sequence of three pulses (5~ms each) resonant with the blue cooling transition at 399~nm and collecting the scattered fluorescence on a photomultiplier tube (PMT).  Each PMT signal is integrated and then digitized by a microprocessor.  The first pulse measures and heats away all atoms remaining in the ground state after spectroscopy; the second pulse measures the background signal due to fluorescence from the thermal atomic beam; and the third pulse measures the atoms that were excited to the long-lived $^3P_0$ state by pumping the atoms back to the ground state with the $^3P_0\rightarrow\!^3D_1$ transition at 1388~nm.  From these three signals (ground state population, excited state population, background level) the microprocessor computes a normalized excitation fraction to reduce the effects of shot-to-shot atom number fluctuations.  We alternately probe the half-maximum points of the resonance from each $m_F=\pm1/2$ ground state, and the microprocessor controls two synthesizers that lock to the average and splitting of the two resonances.  This detection system is nearly projection noise-limited, with a signal-to-noise ratio exceeding 100; however, the clock instability is limited by the clock laser \cite{Ludlow08}.

To evaluate systematic shifts of the clock frequency, we performed a series of optical comparisons against the Ca optical standard at NIST \cite{Oates00} via a self-referenced femtosecond laser frequency comb \cite{Fortier06, Stalnaker07}.  During these measurements, the fractional frequency instability between the two clocks started at 5~$\times~10^{-15}$ at 1~s and averaged down to 3~$\times~10^{-16}$ at 200~s.  On longer time scales, drifts in the Ca clock degrade the stability, so typically data sets were taken in repeated bins of 200~s. 

\begin{figure}
\includegraphics[width=8cm]{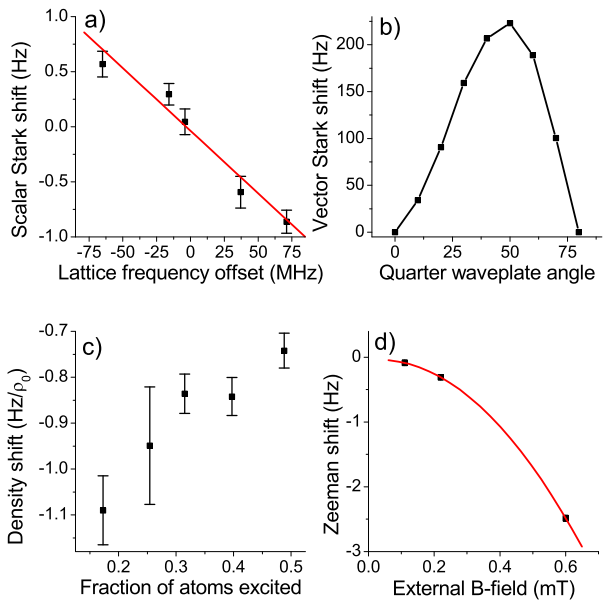}
\caption{Measured systematic effects.  (a) Lattice Stark shift.  By varying the lattice power during spectroscopy for different lattice wavelengths, we find the magic frequency to be 394,798,329(10) MHz.  A linear fit (solid line) gives the slope as -11(1) Hz/GHz.  (b) Vector Stark shift.  By monitoring the splitting between the two $\pi$ resonances as the lattice ellipticity is varied,  we find an upper limit for the vector Stark shift to be 220(30)~Hz for a 50~$\mu$K lattice depth.  (c) Density shift as a function of final excitation, which was varied by changing the lock point on the resonance.  (d) Magnetic field.  Measuring the clock frequency versus external magnetic field, we find the first-order Zeeman to be canceled at the $4~\times~10^{-17}$ level, while the second-order coefficient of -0.07(1)~Hz/G$^2$ is in good agreement with previous work \cite{Poli08,Taichenachev06}. A quadratic fit (solid line) is also shown. \label{systematics}}
\end{figure}

As with all optical lattice clocks, it is necessary to measure the effects of the Stark shift from the confining lattice light.  The first-order effect is canceled by tuning the lattice to the ``magic'' frequency, at which the ground and excited clock states have identical polarizability values \cite{Katori03}.  To find this magic frequency, we varied the lattice intensity to find the Stark shift for a given lattice frequency, and then repeated this procedure for a series of frequencies.  The results of this measurement are shown in Fig.~\ref{systematics}a, and the magic frequency is determined to be 394 798 329(10)~MHz.  This gives a difference between magic frequencies for the two measured isotopes to be $\nu^{174}_{magic}-\nu^{171}_{magic}=$~1146(36)~MHz \cite{Barber08}.  We also note that this measurement is complicated by the density shift induced by a change in the lattice confinement.  The final uncertainty for the scalar Stark shift is given by the measurement statistics at the magic frequency (1.7~$\times~10^{-16}$), which could be substantially reduced by measuring with higher lattice power \cite{Brusch06} to achieve a larger lever arm, with an additional term (1~$\times~10^{-16}$) allowing for possible magnetic dipole and electric quadrupole shifts \cite{Taichenachev08}.

In addition to the first-order scalar Stark shift, there are two other lattice effects that need to be considered.  First, two-photon resonances near the magic frequency lead to hyperpolarizability shifts; these were addressed in \cite{Barber08}.  The shift and uncertainty for our typical lattice depth are shown in Table~\ref{table}.  Second, for odd isotopes, vector and tensor light shifts arise from hyperfine interactions, though for spin-1/2 systems the tensor shift is zero \cite{Angel68}.  As expressed in \cite{Porsev04}, the vector shift for a given $m_F$ state is $\delta f_v=-m_F A \alpha_v \left(\frac{1}{2} \mathcal{E}_0\right)^2$, where $A$ is the degree of circular polarization, $\mathcal{E}_0$ the electric field, and $\alpha_v$ the vector polarizability at 759~nm.  For our experimental geometry, the vector and Zeeman shifts of the $^3P_0$ add in quadrature, allowing the vector shift to be determined by measuring the splitting between $m_F=\pm1/2$ resonances as a function of lattice polarization ellipticity.  As shown in Fig.~\ref{systematics}b, the maximum vector shift (for circular polarization with 50~$\mu$K lattice depth) is $\delta f_v=$220(30)~Hz, where the error bar is derived from imperfect polarization control due to birefringence in the vacuum windows.  This shift corresponds to a vector polarizability of $\alpha_v \approx-$0.08 a.u., in agreement with the calculation in \cite{Porsev04}.  During typical operation, the vector shift is suppressed to less than 10$^{-18}$ by three effects: linear polarization ($A\approx 0$), geometry \cite{Porsev04}, and cancellation by averaging over both $m_F=\pm1/2$ resonances.  For a three-dimensional lattice, the vector shift could be more problematic, as each lattice site could experience a different, time-dependent composite polarization vector.  This can lead to broadening of each resonance by as much as $\delta f_v$, requiring the vector shift to be suppressed by other means (e.g. \cite{Cheng01}).  Even with vector shift-induced broadening, the clock frequency remains unperturbed by proper averaging over both $m_F=\pm1/2$ resonances.

To measure a possible density-dependent collision shift, we lowered the power in the 399~nm slowing beam to reduce the number of atoms loaded into the lattice while looking for a clock shift.  Sideband spectra show that the atom temperature ($\approx 15~\mu$K) is independent of slowing power, ensuring that only the atom number (not the confinement) is varied.  We find a linear relationship between atom number and clock frequency, with a slope of -0.85(4) Hz/$\rho_0$.  We emphasize that while absolute density is difficult to measure, the relative density is well-calibrated by the detection scheme described above.  During typical operation, the excitation fraction is 0.3, at which the density shift uncertainty is 8$\times 10^{-17}$.

We also investigated the density shift dependence on atomic excitation by varying the detuning of the lock points on the clock transition (see Fig.~2c).  Because the magnitude of the shift increased as the atomic excitation was reduced, we first investigated two-body collisions between $^1S_0$ atoms as the dominant shift mechanism.  Founded in the Pauli exclusion principle, as two ground ($^1S_0$) state atoms in the same spin state comprise a pair of identical fermions, s-wave interactions are prohibited.  The atomic temperature is not sufficiently cold to rule out by itself the p-wave term of such an interaction.  However, photoassociation spectroscopy of many Yb isotopes has been used to determine cold collision scattering cross sections of ground state atoms, not only yielding precise s-wave scattering lengths but also enabling calculation of higher wave contributions \cite{Kitagawa08}.  The expected interaction for $^{171}$Yb is not sufficiently large to justify the observed shift.  Furthermore, the interaction is expected to be much larger for $^{174}$Yb, where no collisional shift has been observed at similar densities (albeit with larger uncertainty) \cite{Poli08}.  

Another possibility lies with interactions between $^1S_0$ and $^3P_0$ atoms.  In ultracold fermionic Sr, inhomogeneous excitation of the lattice-confined atoms leads to only partial suppression of s-wave interactions of this type and has been shown to yield a non-zero density shift \cite{Campbell09}.  This same framework, applied to $^{171}$Yb, including only partial suppression by the two body correlation function, appears consistent with our observed shift as a function of excitation.  In contrast to the case of Sr, here no clear zero crossing in the shift is present.  However, based on the above analysis, this zero crossing is expected to occur at higher excitation fraction, and the crossing would be found at reduced excitation for a colder atomic sample.  Colder temperatures would also give overall smaller shifts.  These shifts would be further reduced by sparse population in a higher dimensional lattice.

A summary of all known systematic effects is given in Table \ref{table}.  We determined the Zeeman shifts by measuring changes in the clock frequency as the magnetic field is varied. The resulting data, shown in Fig. 2d, are fit to linear and quadratic functions to determine the first- and second-order sensitivities.  As expected, the first-order effect is canceled by the averaging technique, while the second-order shift is in good agreement with theory \cite{Taichenachev06} and previous measurements \cite{Poli08}.  The probe light shift caused by off-resonant excitations originating from the 578~nm light was calculated based on previous measurements \cite{Poli08} for our typical 20~nW of power.  The acousto-optic modulator (AOM) phase chirp was measured by changing the RF power to the chopping AOM (between cavity and atoms) and looking for a clock shift, as well as by heterodyne interferometry \cite{Degenhardt05}; both of these measurements were consistent with no shift for our typical conditions within the measurement uncertainty.  The blackbody shift for room-temperature operation is calculated based upon the polarizability calculations in \cite{Porsev06}. The 10 $\%$ uncertainty in the polarizability leads to a room-temperature uncertainty of 2.5~$\times~10^{-16}$, and future work to measure the polarizability or operate below room temperature will be necessary to improve the clock's accuracy.  Alternatively, one could define the fundamental clock operating condition at room-temperature, in which case the correction is much smaller and the blackbody uncertainty is 3.3~$\times~10^{-17}$ for 1~K uncertainty in temperature.  Other systematic effects (servo error, residual Doppler shifts) have been calculated to be negligibly small.  

\begin{table}[t] \caption{Uncertainty budget for systematic effects in the $^{171}$Yb optical lattice clock.  Values are given in fractional units $10^{-16}$ and represent typical operating conditions.}
\label{table}
\begin{tabular}{c c c}
\toprule Effect &  Shift& Uncertainty \\
\hline
Blackbody & -25 & 2.5 \\
Lattice polarizability & 0.4 & 2.0\\
Hyperpolarizability &  3.3 & 0.7\\
Density  &  -16.1 & 0.8 \\
First-order Zeeman & 0.4 & 0.4\\
Second-order Zeeman &  -1.67 & 0.1\\
Probe light &  0.05 & 0.2 \\
AOM phase chirp & -0.05 & 0.1 \\
Servo and Line pulling	&	0 & 0.1\\
Residual Doppler & 0 & 0.1\\
\hline
Total &  -38.7 & 3.4\\
\botrule
\end{tabular}
\end{table}

We have completed a comparison of the $^{171}$Yb lattice clock with the NIST-F1 Cs primary standard \cite{Heavner05}.  For these measurements, the femtosecond comb's repetition rate ($\sim$ 1~GHz) was locked to the heterodyne beat with Yb, mixed down to several MHz with a synthesizer referenced to a hydrogen maser, and counted, while the Cs clock simultaneously calibrated the H-maser.  Allan deviation and histogram plots of the measurement statistics give the fractional instability as 5.1~$\times~10^{-13}\tau^{-1/2}$.  After averaging for $>10^5$ seconds, we find the measured frequency is 518~295~836~590~865.2(0.7)~Hz, with the statistical uncertainty (1.4~$\times~10^{-15}$) still much larger than the systematic uncertainties of either clock.  Combined with our previous measurements \cite{Poli08}, we determine the isotope shift $\nu^{171}_{clock}-\nu^{174}_{clock}=$~1 811 281 647.4(1.1)~Hz.

In conclusion, we have completed a systematic evaluation of the $^{171}$Yb optical lattice clock with uncertainty 3.4~$\times~10^{-16}$, competitive with the best neutral-atom microwave and optical frequency standards currently operating.  The simplicity of the spin-1/2 system is enjoyed in several ways, including the simple spectrum, straightforward state polarization, and lack of tensor lattice shifts.  These results indicate that $^{171}$Yb is an excellent candidate not only for an optical frequency standard but also for quantum information experiments that could benefit from a spin-1/2 quantum system with long coherence times and simple nuclear spin manipulation.

The authors thank L.-S. Ma for experimental assistance, S. Bickman and T. Johnson for careful reading of the manuscript, and P. Julienne, S. Blatt, and J. Sherman for helpful discussions.  

\textit{Note added}. --- While completing this manuscript, we became aware of another recent measurement of the $^{171}$Yb clock frequency \cite{Kohno09}, which agrees with our value to within their stated uncertainty.

\end{document}